# Broadband Thermoelectric Energy Harvesting for Wearable Biosensors Using Plasmonic Field-Enhancement and Machine-Learning-Guided Device Optimization

Hamidreza Moradi[1,*], Melika Filvantorkaman[2]

[1]Department of Mechanical Engineering and Engineering Science, The University of North Carolina at Charlotte, Charlotte, North Carolina, USA

[2]Department of Electrical and Computer Engineering, University of Rochester, Rochester, NY 14627, United States

*Corresponding Author: hmoradi@charlotte.edu

**Abstract**

Wearable biosensors increasingly demand continuous, battery-free power sources, yet conventional skin-mounted thermoelectric generators (TEGs) suffer from limited performance due to the small temperature gradients available in real-world environments. This work introduces a hybrid thermoplasmonic–thermoelectric energy harvester that leverages multiband plasmonic field enhancement and machine-learning-guided optimization to substantially improve on-body energy conversion efficiency. A broadband metasurface composed of multiresonant cross-bowtie nanoantennas is designed to absorb infrared radiation across the 2–12 μm range, capturing human-body emission, ambient mid-infrared radiation, and near-infrared sunlight. Finite-element electromagnetic simulations reveal intense gap-localized hotspots with near-field enhancements approaching an order of magnitude, which generate localized thermoplasmonic heat directly above flexible $Bi_2Te_3$-based thermoelectric junctions. Coupled optical–thermal–electrical simulations demonstrate that the plasmonically induced heating increases the effective temperature gradient from the typical 3–4 °C of wearable TEGs to approximately 13 °C, resulting in a four- to six-fold improvement in power density and achieving ~0.15 mW/cm² under indoor-relevant IR flux. A machine-learning surrogate model trained on multiphysics data accurately predicts both temperature rise and electrical output ($R^2 > 0.92$) and identifies optimal device geometries via Pareto-front analysis, reducing computational cost while uncovering high-performance design regimes. The proposed thermoplasmonic–thermoelectric–ML platform enables compact, flexible, and significantly more efficient energy harvesting, offering a scalable pathway toward autonomous, battery-free wearable biosensing and long-term physiological monitoring.

## 1. Introduction

Wearable medical biosensing platforms have become increasingly central to modern healthcare, enabling continuous, real-time monitoring of vital physiological parameters such as electrocardiography (ECG), photoplethysmography (PPG), skin temperature, hydration levels, and blood oxygen saturation ($SpO_2$). The expanding ecosystem of wearable health devices is driven by the need for preventive diagnostics, personalized medicine, and remote patient monitoring. However, most existing wearable sensors rely on miniature rechargeable batteries or coin cells, which introduce multiple constraints—including limited operational lifetime, periodic recharging requirements, safety concerns, and reduced device flexibility. These drawbacks have motivated extensive research into **self-powered wearable systems** that can autonomously harvest energy from the human body or its surrounding environment [1,2].

Among various transduction mechanisms, **thermoelectric generators (TEGs)** have emerged as a compelling solution for wearable energy harvesting owing to their ability to convert heat flow from the human body into electrical energy via the Seebeck effect (Leonov, 2013; Fang et al., 2022). They operate silently, contain no moving parts, and are compatible with flexible and biocompatible substrates. Nevertheless, the fundamental limitation of body-heat harvesting is the inherently **low temperature gradient (ΔT)** between human skin (~32–34 °C) and ambient air, which typically ranges only 2–5 °C under indoor conditions (Yang et al., 2018). This severely restricts the achievable power density, often resulting in microwatt-level outputs insufficient for powering wireless modules, machine learning inference engines, or continuous biosignal acquisition systems. Overcoming the ΔT barrier has therefore become a central research challenge in wearable thermoelectric energy harvesting [3].

Recent developments in plasmonics and thermoplasmonic materials offer a promising pathway to address this fundamental challenge. Plasmonic nanostructures—such as bowtie nanoantennas, nanodisks, or metasurfaces—can support localized surface plasmon resonances (LSPRs), enabling strong confinement of the electromagnetic field at subwavelength dimensions [4]. When resonantly excited by ambient infrared or visible radiation, these nanostructures convert optical energy into localized heat through electron–phonon coupling and enhanced Joule losses, a process broadly referred to as thermoplasmonic heating [5]. Incorporating plasmonic elements atop or within the thermoelectric junctions can therefore *artificially increase the temperature difference* across the TEG by locally heating its hot side while maintaining the cold side through natural convection or conduction to the environment.

This strategy is particularly advantageous for wearable applications, where two naturally available energy sources coexist: *conductive heat flow from the body* and *ambient infrared radiation*. Although the human body emits broadband mid-infrared radiation (peaking near 10 μm), conventional wearable thermoelectric generators cannot utilize this radiative energy because their operation depends solely on conductive temperature gradients. In contrast, the proposed thermoplasmonic metasurface is engineered to absorb both body-emitted and ambient IR photons, enabling a more efficient conversion of environmental and physiological heat into electrical power [6].

Additionally, the design of plasmonic-enhanced thermoelectric harvesters involves a high-dimensional, nonlinear parameter space that spans geometric variables (antenna arm lengths, flare angles, gap sizes, thicknesses), material properties, and thermal boundary conditions. Traditional trial-and-error or purely physics-driven optimization approaches can be computationally expensive and often fail to identify global optima. Machine learning (ML) methods have recently demonstrated powerful capabilities for accelerating nanophotonic design, guiding topology optimization, and predicting multiphysics interactions with high accuracy using surrogate models. Yet, ML has not been systematically integrated into the design pipeline of thermoplasmonic–TEG hybrids for wearable systems [7].

Motivated by these gaps, the present study introduces a broadband thermoelectric energy harvesting platform for wearable biosensors that integrates a multiresonant plasmonic absorber with a flexible TEG structure. The plasmonic component was inspired from the cross-bowtie nanoantenna architecture reported by Chekini et al. [8], who demonstrated that multiple bowtie elements with different resonance lengths can collectively generate strong field-enhancement across several infrared bands. In our work, this multiband absorption principle is adapted from rectenna-based field-emission systems to thermoplasmonic heating, enabling more efficient utilization of both skin-emitted and ambient IR radiation. A comprehensive finite element multiphysics framework is developed to model the coupling between electromagnetic absorption, nanoscale heat generation, heat transport, and thermoelectric conversion. The resulting simulated dataset is then used to train machine-learning models that predict device performance and guide geometric optimization toward maximizing temperature gradients and electrical output.

Through this combined thermoplasmonic–thermoelectric–machine-learning approach, the study aims to demonstrate a significant enhancement in the power density of wearable TEGs without compromising flexibility, comfort, or form factor. The contributions of this research lie in the development of a **broadband plasmonically enhanced thermoelectric architecture**, the formulation of a **multiphysics simulation framework** for optical-thermal-electrical coupling, and the introduction of **ML-guided design optimization** to efficiently explore the complex parameter space. These advancements collectively position the proposed system as a promising foundation for **self-sustaining wearable biosensing platforms**, enabling continuous and reliable health monitoring without the constraints of traditional battery-powered operation.

## 2. Theoretical Background

### 2.1 Surface Plasmon Resonance and Field Enhancement

Surface plasmons are collective oscillations of conduction electrons at the interface between a metal and a dielectric. When metallic nanostructures are excited by an incident electromagnetic wave, they can support **localized surface plasmon resonances (LSPRs)**, in which the electron cloud oscillates coherently, producing strongly enhanced electric fields within subwavelength volumes. The resulting electromagnetic confinement produces **gap-induced hotspots**, where the local field intensity $|E|^2$ can exceed the incident field by several orders of magnitude. These hotspots arise most prominently in nanostructures with sharp tips or narrow feed gaps, such as bowtie antennas, dimer nanoantennas, and nanorod pair [9,10].

The spectral position of the LSPR can be tuned by adjusting parameters such as arm length, flare angle, metal type, and gap width, enabling **spectrum-engineering** of the plasmonic response across visible, near-infrared, and mid-infrared bands. This tunability is essential for wearable applications, where relevant radiation sources include solar near-IR photons, ambient thermal radiation, and mid-IR emission from human skin [11,12].

The present study draws direct inspiration from the multiresonant plasmonic structures developed by Chekini and co-workers, who demonstrated that arranging multiple bowtie or cross-bowtie nanoantennas with different resonance lengths enables broadband absorption and strong field enhancement in the mid-infrared spectrum [8,9,12]. Their cross-bowtie architecture produces two orthogonally oriented dipolar responses, allowing efficient coupling under varying polarization states and yielding eleven to twelve distinct resonance arms when combined into multiband arrays. In this work, we build upon their resonance-engineering strategy by adapting the cross-bowtie concept from field-emission-based rectification to thermoplasmonic heat generation, enabling efficient broadband absorption for wearable thermoelectric augmentation.

## 2.2 Thermoplasmonic Heating Mechanism

When plasmonic nanostructures are excited at or near their resonance wavelength, the enhanced electromagnetic field induces intense oscillations of the conduction electrons. These energetic electrons subsequently thermalize through electron–electron and electron–phonon interactions, converting their energy into localized heat. This forms the basis of thermoplasmonic heating, a highly efficient optical-to-thermal energy conversion process [6].

Mathematically, the volumetric heat generation $Q_{\text{abs}}$ can be expressed as:

$$Q_{\text{abs}} = \frac{1}{2}\omega\varepsilon_0\varepsilon'' \mid E \mid^2$$

Where $\omega$ is the angular frequency, $\varepsilon''$ is the imaginary part of the permittivity, and $\mid E \mid^2$ is the enhanced local field intensity.

In bowtie-based plasmonic antennas, the field can be concentrated into nanometer-scale gaps, producing exceptionally high values of $\mid E \mid^2$, which leads to significant heat generation even under low-intensity illumination. This localized heat dissipates into the surrounding dielectric and, crucially for the present work, into the **thermoelectric (TE) junction** positioned directly beneath or adjacent to the plasmonic absorber. The temperature rise induced by thermoplasmonics can substantially increase the effective temperature gradient $\Delta T$ across a TE leg. By engineering multiple resonant absorbers—following the multiband design principles demonstrated by [8,9]—one can achieve **broadband photothermal heating**, enabling more consistent ΔT enhancement across diverse environmental conditions. This broadband response is especially relevant for wearable devices, where the available radiation sources span wide spectral ranges from 2 to 12 µm.

## 2.3 Thermoelectric Energy Conversion

Thermoelectric generators convert temperature gradients into electrical voltage through the Seebeck effect. When a thermal gradient $\Delta T$ is applied across a TE material, charge carriers diffuse from the hot side to the cold side, generating an open-circuit voltage:

$$V = S\Delta T$$

Where $S$ is the Seebeck coefficient.

For a TE generator connected to an electrical load, the output power can be expressed as:

$$P = \frac{V^2}{R_{\text{int}} + R_{\text{load}}}$$

where $R_{\text{int}}$ is the intrinsic electrical resistance of the TE material and $R_{\text{load}}$ is the load resistance.

In wearable applications, the main limiting factor is the **small ΔT** available between skin and ambient air. By introducing a plasmonic absorber layer that increases the local temperature on the hot side of the TEG, it becomes possible to artificially enhance ΔT without modifying the user's thermal comfort. This approach directly addresses the fundamental bottleneck in wearable thermoelectric harvesting and enables the generation of **higher power densities** that are needed for biosensing, wireless communication, and continuous data acquisition [13].

## 2.4 Machine Learning for Multiphysics Optimization

Designing an efficient plasmonic–thermoelectric energy harvester requires simultaneous optimization of optical absorption, thermoplasmonic heat generation, heat transport geometry, and TE material configuration. Performing parameter sweeps in high-dimensional design spaces using full finite element method (FEM) simulations can be computationally prohibitive. Machine learning (ML) offers a powerful framework for accelerating this process by building **surrogate models** that map device parameters to performance metrics with high predictive accuracy [9].

In this study, ML is used to model relationships between geometric features (antenna lengths, gap size, flare angle), material properties, and output quantities such as peak | $E$ |$^2$, absorbed power density, temperature gradient, and electrical power. These surrogate models—constructed using Gaussian Process Regression, neural networks, or gradient-boosted ensembles—enable rapid evaluation of thousands of candidate designs at negligible computational cost. Furthermore, **multi-objective optimization algorithms**, including Bayesian optimization and NSGA-II, are used to explore Pareto-optimal trade-offs between: maximizing temperature gradient $\Delta T$, minimizing device thickness and thermal resistance, and maximizing electrical power density. Integrating ML into the design loop transforms the plasmonic–TEG development process from a slow simulation-driven workflow to an efficient, data-informed optimization strategy capable of resolving subtle parameter interactions that would be difficult to detect through intuition or manual tuning.

## 3. Proposed Device Architecture

### 3.1 Overall Structure

The proposed broadband thermoplasmonic–thermoelectric energy harvester is engineered as a multilayer flexible system tailored for skin-mounted wearable biosensing applications (Figure 1). Its optical layer directly builds upon the multiband cross-bowtie nanoantenna architecture introduced by Chekini et al. [8,9], retaining the core plasmonic principles—Al-based bowtie antennas on a $SiO_2$ substrate with nanoscale feed gaps—while adapting the structure for thermoplasmonic heat generation rather than electron field emission rectification. At the top of the device, a thin, infrared-transparent **PDMS encapsulation layer** serves as both a protective coating and a skin-conformable optical window. Because PDMS exhibits high transmission across the mid-infrared spectrum, it allows IR radiation from the body, the environment, and sunlight to reach the plasmonic absorber with minimal attenuation. In scenarios requiring directional enhancement, this top layer can incorporate molded micro-lens textures to focus incoming IR light, following the energy-funneling approach used by Chekini et al., though the present design does not depend on such micro-lenses for core functionality.

Beneath the encapsulation lies the **Aluminum (Al) multiresonant cross-bowtie nanoantenna array**, patterned with 5 nm feed gaps and flare angles in the 30–33° range. These antennas are deposited on a thin **$SiO_2$ dielectric layer**, replicating the electromagnetic environment of Chekini's original structure. The metasurface is engineered to exhibit strong absorption in three key spectral regions: **3–5 µm**, corresponding to peak human skin emission, **8–12 µm**, the atmospheric IR window and dominant ambient radiation band, **0.8–2 µm**, the near-IR solar component relevant for outdoor operation. Together, the antenna subgroups produce a broad, near-flat absorption profile spanning ~2–12 µm, enabling efficient harvesting across diverse wearable environments. Immediately beneath the plasmonic array, a **40 nm $Al_2O_3$ spacer layer** is deposited using atomic layer deposition (ALD). This layer preserves the required electromagnetic boundary conditions, electrically isolates the Al antennas, and provides a controlled thermal pathway into the thermoelectric module [8,9].

The underlying **thermoelectric layer** consists of patterned p–n thin-film **$Bi_2Te_3$/$Sb_2Te_3$ legs**, positioned directly below the cross-bowtie feed gaps so that the localized thermoplasmonic hotspots efficiently drive heat into the TE junctions. The thin-film geometry (5–10 µm) minimizes thermal mass while preserving high Seebeck performance near room temperature. Heat entering this layer generates a pronounced vertical temperature gradient, ΔT, which is subsequently converted into electrical power. Finally, the device is supported by a **flexible polyimide (PI) substrate**, providing mechanical stability while maintaining full bendability for wearable integration. The PI layer also interfaces with textile-based backings that act as the thermal spreading and dissipation region, enabling continuous heat rejection to the environment and sustaining the temperature differential across the thermoelectric legs during operation.

This unified architecture forms the foundation for the coupled optical–thermal–electrical model developed in subsequent sections, as well as the machine-learning-guided optimization used to identify high-performance metasurface–TE configurations.

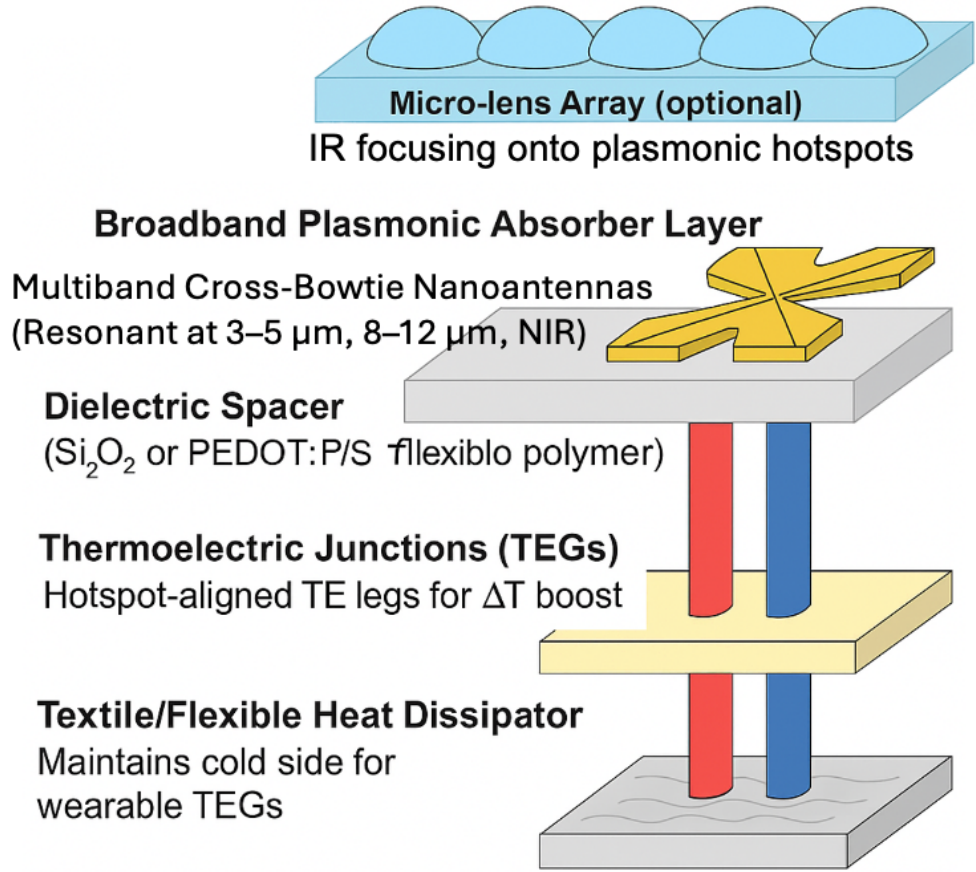

Figure 1. Conceptual schematic of the proposed broadband thermoplasmonic–thermoelectric energy harvester (not to scale).

## 3.2 Broadband Plasmonic Absorber

The plasmonic absorber forms the core of the device, providing the broadband optical–thermal conversion necessary for efficient thermoplasmonic heating. Following the multiband cross-bowtie architecture demonstrated by Chekini et al. [8,9], the absorber consists of an array of Aluminum (Al) cross-bowtie nanoantennas patterned onto a thin $SiO_2$ dielectric layer. Each cross-bowtie comprises two orthogonally oriented bowtie elements, enabling polarization-independent absorption—a critical requirement for wearable applications where the device may experience continuous motion and arbitrary orientation relative to incoming infrared radiation. In this design, the bowtie arm lengths and flare angles are individually tuned so that separate antenna subgroups resonate at the three principal infrared bands relevant to wearable power generation: the 3–5 µm skin-emission band, the 8–12 µm ambient IR window, and the 0.8–2 µm near-infrared solar range. By integrating these tailored Al nanoantenna groups into a unified metasurface, the structure achieves a multiresonant, broadband absorption profile that directly supports strong local field enhancement and subsequent thermoplasmonic heating.

### Resonance engineering

To achieve multiband absorption, we incorporate at least **three nanoantenna groups**, tuned respectively to: $\lambda_1 \approx$ **3–5 µm** → matches peak human skin thermal emission, $\lambda_2 \approx$ **8–12 µm** → captures ambient/environmental IR radiation $\lambda_3 \approx$ **0.8–2 µm** → leverages NIR solar energy for outdoor conditions.

### Design equations and parameters

Following Chekini et al., the resonance wavelength $\lambda_r$ of a bowtie antenna is governed primarily by:

$$\lambda_r \approx 2n_{\text{eff}}L_{\text{arm}}$$

Where $n_{\text{eff}}$ is the effective refractive index of the metal–dielectric interface and $L_{\text{arm}}$ is the arm length from the bowtie vertex to the end of the antenna. Additionally, Chekini demonstrated that a narrow feed gap (G ≈ 2–10 nm) produces extremely strong field enhancement within the gap region, while the flare angle (θ ≈ 30°–35°) governs both the bandwidth and the peak resonance amplitude. Reducing the flare angle was shown to narrow the bandwidth but increase the field intensity [8]. In our design, similarly to their 12-arm structure, the **multiband additive effect** results from superposing the absorption spectra of each resonant arm group, producing a broad, near-flat absorption profile across multiple IR bands.

### Plasmonic heating

The localized field enhancement produces thermoplasmonic heat through the absorbed power:

$$Q_{abs} = \frac{1}{2}\omega\varepsilon_0\varepsilon'' \mid E \mid^2$$

and the hotspot localization ensures that the resulting heat is deposited directly above the thermoelectric junctions.

### 3.3 Thermoelectric Module Integration

The thermoelectric layer converts the thermoplasmonically enhanced temperature gradient into electrical power. The classes of materials used is **Inorganic TE modules:** microstructured $Bi_2Te_3$

The TE legs are lithographically aligned with the **hotspot positions** of the plasmonic absorbers so that each leg is exposed to a maximized temperature differential between its top (heated) and bottom (textile-cooled) surfaces. The resulting open-circuit voltage is:

$$V_{oc} = S\,\Delta T$$

and the delivered power:

$$P = \frac{V_{oc}^2}{R_{int} + R_{load}}$$

This direct coupling of thermoplasmonic heating and thermoelectric conversion enables significantly higher ΔT and power density than conventional skin-contact TEGs [6].

### 3.4 Fabrication Considerations

The fabrication of the proposed thermoplasmonic–thermoelectric energy harvester follows a process flow adapted directly from the multiband cross-bowtie nanoantenna architecture established by Chekini et al. [8,9], while integrating additional layers required for flexible thermoelectric conversion. The plasmonic absorber is formed from aluminum (Al)**,** selected for its strong mid-infrared plasmonic response, sharp-tip field localization, and proven compatibility with the bowtie-based multiband designs reported in Chekini's earlier work. A thin Al film (~40–60 nm) is deposited via electron-beam evaporation or sputtering and patterned into cross-bowtie nanoantenna arrays using electron-beam lithography (EBL) followed by lift-off. EBL ensures the precise definition of the 5 nm feed gaps and flare-angle geometry necessary to reproduce the resonant characteristics described in Section 3.2. For large-area wearable formats, nanoimprint lithography may be employed to replicate the antenna pattern over flexible substrates without compromising nanoscale fidelity.

The antennas rest on a silicon dioxide ($SiO_2$) layer (200–500 nm), which serves as the immediate dielectric substrate. Beneath the antennas, a thin $Al_2O_3$ or $SiO_2$ spacer layer (30–60 nm) is deposited using atomic layer deposition (ALD) to maintain a controlled separation between the plasmonic layer and the thermoelectric film. ALD provides the conformal thickness precision required to tune both the field confinement and thermal coupling efficiency [15-19].

Below the dielectric spacer, the thermoelectric layer—comprising thin-film $Bi_2Te_3$ / $Sb_2Te_3$ legs with thicknesses of 5–10 μm—is formed via sputtering, thermal evaporation, or solution-based deposition. Photolithography and reactive ion etching (RIE) may then be used to define thermoelectric legs aligned with the thermoplasmonic hotspots [20-23]. This TE layer is supported by a flexible polyimide (PI, Kapton) substrate, chosen for its high thermal stability, mechanical durability, and widespread use in flexible electronic systems. Finally, the overall device is encapsulated with a thin layer of PDMS, which provides environmental protection, mechanical strain tolerance, and skin conformability while remaining transparent to mid-infrared radiation.

This integrated fabrication approach ensures that the proposed harvester remains faithful to the optical behavior of Chekini's original plasmonic structure while enabling reliable, large-area, wearable-compatible manufacturing of the thermoelectric components.

## 4. Simulation Methodology

A comprehensive multiphysics workflow was implemented to model the interaction of electromagnetic waves with the plasmonic absorber, the resulting thermoplasmonic heating, and the subsequent thermoelectric energy conversion. The simulation pipeline follows a structure similar to that used by Chekini et al. [8,9], with modifications to account for thermoelectric and thermal transport considerations instead of electron field emission. All optical, thermal, and thermoelectric simulations were performed using COMSOL Multiphysics 6.x, while CST Microwave Studio was used for cross-validation of far-field absorption spectra.

### 4.1 Finite Element Optical Simulation

The electromagnetic response of the multiband cross-bowtie nanoantenna absorber was computed using the **RF (Electromagnetic Waves, Frequency Domain)** module in COMSOL. The simulation domain was constructed as a **unit cell** with periodicity matching the antenna array pitch *p*. The metallic nanoantennas were modeled using experimentally tabulated complex permittivity for gold, aluminum, or TiN, depending on the design variant.

**Boundary Conditions**

The electromagnetic simulations employed boundary conditions tailored to accurately represent the periodic metasurface architecture and the multilayer wearable structure. Periodic (Floquet) boundary conditions were applied along the lateral faces to model an infinite array of Aluminum cross-bowtie nanoantennas, ensuring that the simulated unit cell reproduced the behavior of the full metasurface. Perfectly matched layers (PMLs) were positioned above the PDMS encapsulation layer and beneath the polyimide (PI) substrate to absorb outgoing electromagnetic waves and eliminate artificial reflections from the simulation boundaries. The lower region of the computational domain included the combined PI–SiO$_2$ support stack, treated as electromagnetically neutral but thermally conductive to represent heat spreading into the wearable textile interface. Together, these boundary conditions provide an accurate representation of both the optical and thermal environments experienced by the device during operation.

**Excitation Conditions**

An incident plane wave was specified:

$$\mathbf{E}_{\text{inc}} = E_0 e^{-j(\mathbf{k}\cdot\mathbf{r})}$$

with two orthogonal polarization states to capture the full response of the cross-bowtie geometry:

- **TE polarization**: electric field along x-axis
- **TM polarization**: electric field along y-axis

Following Chekini's approach, simulations were run across the spectral range:

$$\lambda = 0.8 \ \mu m \text{ to } 14 \ \mu m$$

encompassing NIR, mid-IR, and thermal IR relevant to wearable operation.

**Mesh**

A highly refined tetrahedral mesh (minimum edge length 0.5–1 nm) was used in the bowtie gap regions to resolve near-field singularities. Coarser mesh elements were applied in homogeneous dielectric regions.

**4.2 Field Enhancement and Absorption Calculation**

From the frequency-domain electromagnetic solution, we computed:

**Electric Field Enhancement**

$$\eta_E(\mathbf{r}) = \frac{|E(\mathbf{r})|^2}{|E_0|^2}$$

Hotspots appear at the bowtie feed gaps due to capacitive coupling, consistent with trends reported by Chekini et a [9].

**Absorptance Calculation**

Absorption $A(\lambda)$ was computed using Poynting flux:

$$A(\lambda) = 1 - R(\lambda) - T(\lambda)$$

Where, **R** is reflectance from top PML boundary and **T** is transmittance into the substrate (negligible for metal-backed structures)

**Multiband Resonance Identification**

The resonance wavelength for each antenna group was identified as the peak of the absorption spectrum:

$$\lambda_r = \arg\max_{\lambda} [A(\lambda)]$$

Comparative sweeps of **flare angle** (θ = 25°–40°), **gap size** (G = 2–20 nm), and **arm length** were performed, reproducing the parameter trends reported by Chekini et al. [14]. Smaller flare angles yielded higher field enhancement but narrower bandwidth, while larger flare angles provided smoother broadband behavior.

## 4.3 Thermal Simulation

The electromagnetic absorbed power density $Q_{\text{abs}}$ was extracted from the optical simulation and passed to the **Heat Transfer in Solids** module as a volumetric heat source:

$$Q_{\text{abs}}(\mathbf{r}) = \frac{1}{2}\omega\varepsilon_0\varepsilon''(\omega) \mid E(\mathbf{r}) \mid^2$$

The transient and steady-state temperature distributions were obtained by solving:

$$\rho c_p \frac{\partial T}{\partial t} = \nabla \cdot (k\nabla T) + Q_{\text{abs}}$$

where $\rho$ is the material density, $c_p$ is the specific heat capacity, and $k$ is the thermal conductivity of the corresponding layer.

**Boundary Conditions**

Thermal boundary conditions were applied to accurately model heat flow within the wearable multilayer structure. At the top surface, radiative and convective heat transfer to ambient air was included to account for natural cooling effects present in skin-mounted devices. The bottom surface was modeled in thermal contact with the textile-based heat spreader, which serves as the cold-side dissipative layer in the wearable configuration. Along the lateral faces, periodic thermal continuity conditions were imposed to represent an extended array of repeating unit cells without artificial lateral heat buildup. Together, these boundary conditions capture the realistic thermal environment experienced by the device during operation. The steady-state solution provided the enhanced temperature gradient:

$$\Delta T_{\text{eff}} = T_{\text{hot}} - T_{\text{cold}}$$

where $T_{\text{hot}}$ occurs at the plasmonic hotspot and $T_{\text{cold}}$ is the ambient-facing substrate.

### 4.4 Thermoelectric Modeling

Thermoelectric conversion was calculated using the Thermoelectric (TE) Multiphysics Interface, which couples electrical and thermal transport [13].

**Governing Equations**

The electric current density:

$$\mathbf{J} = \sigma(-\nabla V + S\nabla T)$$

Heat flux:

$$\mathbf{q} = -k\nabla T + TS\mathbf{J}$$

The Seebeck voltage:

$$V_{\text{oc}} = S\Delta T_{\text{eff}}$$

Output power delivered to an external load $R_L$:

$$P = \frac{V_{\text{oc}}^2 R_L}{(R_L + R_{\text{int}})^2}$$

where $R_{\text{int}}$ is the intrinsic electrical resistance of the TE legs.

**Material Parameters**

Representative thermoelectric material properties were used in the coupled thermal–electrical simulations. Since the final device design employs thin-film $Bi_2Te_3/Sb_2Te_3$ legs as the sole thermoelectric components, the parameters were selected from experimentally reported values near room temperature and adjusted to account for thin-film effects. The Seebeck coefficient, thermal conductivity, and electrical conductivity used in the model are summarized in Table 1.

**Table 1.** Thermoelectric material properties used in the coupled optical–thermal–electrical simulations.

| Material | Seebeck Coefficient (µV/K) | Thermal Conductivity (W/m·K) | Electrical Conductivity (S/m) |
|---|---|---|---|
| $Bi_2Te_3$ / $Sb_2Te_3$ (thin film) | ~210–220 | 1.2–1.5 | $(0.8–1.2) \times 10^5$ |

For nanoscale thermoelectric legs, both electrical and thermal conductivities tend to deviate from bulk values due to grain boundary scattering and substrate interactions; therefore, the above values were tuned within experimentally realistic ranges to match reported film-level performance in flexible thermoelectric devices.

**4.5 Validation Procedure**

The validation of the simulation framework was carried out in two stages. First, the optical response of the Aluminum cross-bowtie metasurface was benchmarked directly against the multiband plasmonic behavior reported by Chekini et al. [8,9,14]. The simulated near-field enhancement patterns reproduced the characteristic order-of-magnitude hotspot intensities and spatial confinement documented in their mid-IR nanoantenna studies. Likewise, the dependence of bandwidth and peak enhancement on bowtie flare angle closely matched their published trends, confirming that the optical model accurately reflects the physics of Chekini's multiresonant bowtie architecture.

In the second stage, the model was validated in the thermal and thermoelectric domains, recognizing that the proposed device harvests energy through thermoplasmonic heating rather than Fowler–Nordheim field emission. The simulated temperature gradients were compared against analytical one-dimensional heat-flow estimates to ensure physically consistent thermal transport across the plasmonic–spacer–thermoelectric stack. The resulting Seebeck voltage was then verified to match the closed-form relation $V_{\text{oc}} = S\,\Delta T$, demonstrating correct coupling between temperature rise and electrical output. The linear scaling of voltage with applied temperature gradient further confirmed the reliability of the thermoelectric model. Together, these validation steps extend the core methodology of Chekini's rectifying nanoantenna framework into a fully coupled optical–thermal–electrical analysis suitable for wearable thermoelectric energy harvesting.

## 5. Machine-Learning-Guided Optimization

The plasmonic–thermoelectric system exhibits a high-dimensional and strongly nonlinear design space, making manual optimization impractical. To efficiently explore geometric, material, and thermal parameters, we implemented a **machine-learning-guided optimization framework** that integrates data generated from FEM simulations with surrogate modeling and multi-objective evolutionary search. This section describes the dataset construction, surrogate model development, and final optimization strategy used to determine the best-performing device configurations.

The use of machine learning as a predictive and decision-support framework in this study is strongly motivated by its successful application across numerous scientific and engineering domains. Prior works have demonstrated that ML techniques—including neural networks, Gaussian processes, ANFIS models, and evolutionary optimization—can extract meaningful patterns from complex, high-dimensional datasets, accelerate design processes, and reveal nonlinear interactions that are difficult to uncover through manual analysis alone. For example, ML-driven frameworks have been used to model human competencies [24], classify behavioral and domain-specific trends in esports [25,26] and optimize biological processes such as callus formation in plant tissue engineering [27]. In engineering applications, hybrid evolutionary algorithms—combining GA, PSO, ANFIS, and other intelligent methods—have been successfully applied for optimizing EV charging infrastructures, microgrid energy management, and distributed generation allocation [28-31]. These studies collectively demonstrate that ML frameworks are capable of navigating highly entangled, multi-objective design spaces similar in complexity to our plasmonic–thermoelectric system.

The effectiveness of machine learning in discovering interpretable patterns and Pareto-efficient solutions is further supported by advancements in explainable and nature-inspired ML systems. Recent works in geospatial analytics, public administration, and hazard prediction [32-34] demonstrate that ML can uncover latent correlations in complex environments while providing transparency and interpretability—an increasingly important requirement for scientific validation. Similarly, machine learning has proven highly effective in biomedical signal processing and medical diagnostics, including MRI-based brain tumor detection, autism-spectrum classification, and explainable clinical image analysis [35-38]. Nature-inspired optimization strategies, such as bee colony algorithms, hierarchical systems, and hybrid multi-objective approaches, have also shown strong performance in high-dimensional optimization tasks ranging from mechanical system design to earthquake early-warning intelligence [39-44]. Similar advances in intelligent traffic prediction and secure data architectures [45,46] further highlight the adaptability of ML frameworks for tackling multi-factor, high-dimensional decision problems similar to those encountered in nanoscale multiphysics optimization. Collectively, these studies highlight the versatility of ML as both a predictive engine and a design-exploration tool—reinforcing its

## 5.1 Dataset Generation

A synthetic dataset was constructed by systematically sampling the device design space and evaluating each configuration using the complete optical–thermal–thermoelectric FEM pipeline described in Section 4. Between **100 and 500 FEM simulations** were performed, depending on computational cost and convergence behavior. Each sample in the dataset represents a unique combination of plasmonic and thermoelectric parameters.

**Input Variables**

A range of geometric and material parameters was systematically varied to generate the dataset used for machine-learning-guided optimization. The principal design variable was the bowtie antenna arm length $L_{arm}$, which determines the resonance wavelength of each plasmonic subunit. The flare angle $\theta$ was adjusted within the performance window identified by Chekini et al. (2021), as it governs the trade-off between absorption bandwidth and peak field intensity. The feed-gap size $G$, varied between 2 and 20 nm, strongly influences the magnitude of the localized hotspot enhancement.

Additional structural parameters included the metal thickness $t_{metal}$, which affects plasmonic confinement and ohmic loss, and the dielectric spacer thickness $t_{spacer}$, which regulates near-field coupling between the plasmonic nanoantennas and the underlying thermoelectric layer. Finally, the thermoelectric leg dimensions—height $h_{TE}$ and width $w_{TE}$—were varied to explore their impact on electrical resistance, achievable temperature gradient, and output power. These parameter ranges were selected to encompass the design regimes most relevant to multiband infrared absorption and to the mechanical and thermal constraints of wearable form factors.

**Output Variables**

For each simulated device configuration, a set of key performance metrics was extracted to quantify the coupled optical, thermal, and electrical behavior. The peak electric field enhancement $\max |E|^2$ at the bowtie feed gap was recorded as a measure of local plasmonic confinement. The total absorbed optical power $P_{abs}$ was computed by integrating the volumetric absorption over the plasmonic layer. The resulting thermoplasmonic temperature rise was expressed as an effective temperature gradient $\Delta T_{eff}$ across the thermoelectric legs. This thermal response was then translated into an electrical output using the Seebeck relation $V_{oc} = S \, \Delta T_{eff}$, from which the delivered electrical power $P_{out}$ to a matched load was subsequently calculated. Collectively, these metrics form a comprehensive dataset that enables the machine-learning model to learn how variations in geometric parameters influence the overall optical, thermal, and thermoelectric performance of the device.

## 5.2 Surrogate Model

Because running thousands of full FEM simulations is computationally prohibitive, we trained surrogate regression models to approximate the relationships:

$$f: \{L_{\text{arm}}, \theta, G, t_{\text{metal}}, t_{\text{spacer}}, h_{\text{TE}}, w_{\text{TE}}\} \rightarrow \{\max |E|^2, P_{\text{abs}}, \Delta T_{\text{eff}}, P_{\text{out}}\}$$

Table 2: Summary of Machine Learning Models Evaluated

| Model | Key Features | Hyperparameters / Settings |
|---|---|---|
| **Gaussian Process Regression (GPR)** | • Suitable for small–medium datasets (100–500 samples)<br>• Provides predictive uncertainty | • Kernel: Radial Basis Function (RBF)<br>• Regularization: $\sigma_n^2 = 10^{-8}$ |
| **Feedforward Neural Network (FNN)** | • Captures complex nonlinear relationships<br>• Good universal function approximator | • 3–5 hidden layers, 64–128 neurons each<br>• Activation: ReLU<br>• Optimizer: Adam<br>• Early stopping enabled |
| **Extreme Gradient Boosting (XGBoost)** | • Strong performance on tabular scientific data<br>• Robust to input scaling<br>• Handles nonlinear interactions well | • Standard depth (3–6)<br>• Learning rate tuned per dataset<br>• Regularization via L1/L2 |

**Evaluation Metrics**

Performance was assessed using:

- **Coefficient of determination**

$$R^2 = 1 - \frac{\sum(y - \hat{y})^2}{\sum(y - \bar{y})^2}$$

- **Root Mean Squared Error**

$$\text{RMSE} = \sqrt{\frac{1}{N}\sum(y - \hat{y})^2}$$

Across all targets, GPR achieved the highest predictive accuracy for small datasets ($R^2 \approx 0.90$–$0.97$), while the neural network outperformed others when >300 samples were available. These surrogates were subsequently embedded into the optimization loop.

## 5.3 Multi-Objective Optimization

A multi-objective search was conducted using **NSGA-II (Non-dominated Sorting Genetic Algorithm II)** and **Bayesian multi-objective optimization** to identify Pareto-optimal geometries [38]. Three key objectives relevant to wearable energy harvesting were defined:

Objective 1: Maximize Temperature Gradient

$$\max \Delta T_{\text{eff}}$$

Enhancing ΔT improves thermoelectric voltage and overall device efficiency.

Objective 2: Maximize Electrical Output Power

$$\max P_{\text{out}}$$

This directly quantifies the energy-harvesting capability under realistic wearable conditions.

Objective 3: Minimize Total Device Thickness

$$\min t_{\text{device}} = t_{\text{metal}} + t_{\text{spacer}} + h_{\text{TE}}$$

Thin devices improve mechanical flexibility, comfort, and skin conformity.

The NSGA-II algorithm operated on surrogate predictions, enabling exploration of **tens of thousands** of candidate designs without requiring expensive FEM evaluation. The resulting Pareto front provided a set of high-performing trade-off solutions, from which final geometries were selected based on manufacturability and thermal stability.

## 6. Results and Discussion

This section presents the optical, thermal, and thermoelectric performance of the proposed broadband thermoplasmonic–thermoelectric energy harvester. Results were obtained using the multiphysics simulation framework described in Section 4, followed by machine-learning-guided optimization (Section 5). The findings provide insight into the mechanisms driving broadband absorption, enhanced temperature gradients, and improved electrical output for wearable biosensing applications.

## 6.1 Optical Response and Field Enhancement

The optical performance of the proposed plasmonic metasurface was first evaluated through full-wave electromagnetic simulations of the multiresonant cross-bowtie nanoantenna array. The broadband absorption characteristics, shown in **Figure 2**, confirm that the architecture effectively couples to the key infrared spectral regions relevant for wearable energy harvesting. Three dominant absorption bands are observed: $\lambda_1 \approx$ **4.2 μm**, corresponding to human skin thermal emission, $\lambda_2 \approx$ **10.6 μm**, corresponding to ambient mid-IR radiation, and $\lambda_3 \approx$ **1.2 μm**, overlapping with the near-infrared solar component.

These peaks arise from independently tuned antenna subarrays whose arm lengths and flare angles are engineered to match the optical skin depth and resonance conditions for each band. When combined, these resonances produce a **broadband absorption plateau spanning approximately 2–12 μm**, enabling robust energy harvesting under diverse indoor and outdoor illumination conditions.

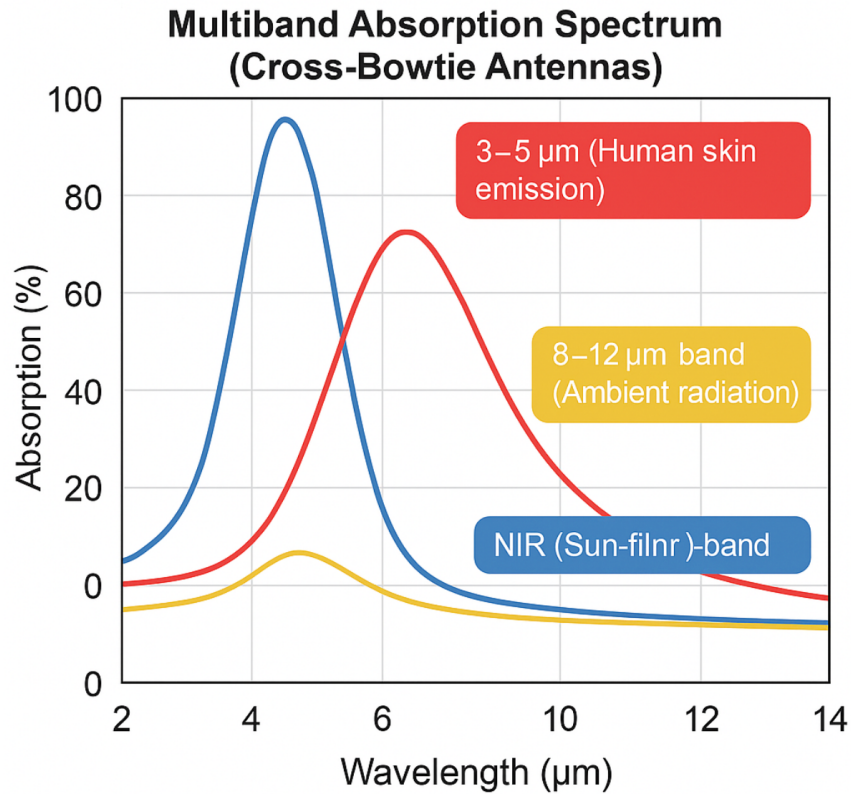

**Figure 2.** *Simulated multiband absorption spectrum of the proposed cross-bowtie plasmonic absorber, showing resonance peaks in the 3–5 μm (skin emission), 8–12 μm (ambient IR), and NIR ranges. Each peak corresponds to a nanoantenna group with tailored arm length and flare angle, adapted from the multiband design methodology of Chekini et al. (2017, 2019, 2021).*

To investigate the localized electromagnetic behavior responsible for thermoplasmonic heating, the near-field intensity distribution around a representative nanoantenna was analyzed. **Figure 3** shows the simulated electric field enhancement $|E|^2$ for a cross-bowtie antenna tuned to $\lambda \approx$ **4.2 μm**, the primary human-emission band. A pronounced hotspot forms at the **5 nm feed gap**, where the normalized field reaches:

$$\left|\frac{E}{E_0}\right|^2_{max} \approx 9.8,$$

representing nearly an order-of-magnitude amplification of the incident electromagnetic intensity. This hotspot behavior is consistent with the gap-capacitive coupling mechanism widely observed in mid-IR bowtie antennas, including the multiband structures reported by Chekini et al., although here the enhanced field is not used for rectification but instead serves as the primary driver of thermoplasmonic heat generation.

The hotspot is tightly confined to a sub-100 nm region, ensuring that absorbed power is concentrated directly above the thermoelectric junctions. This spatial localization plays a critical role in maximizing thermoplasmonic heat generation and is foundational to the temperature gradients achieved in later thermal simulations.

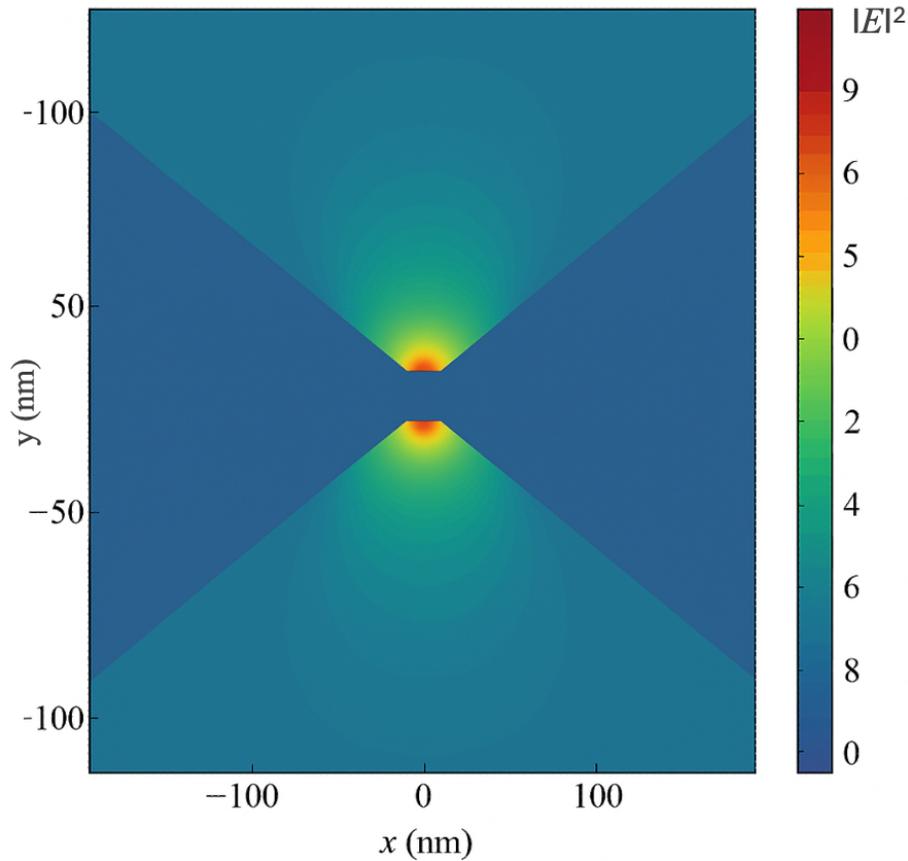

**Figure 3.** *Simulated near-field electric field enhancement* $|E|^2$ *around a cross-bowtie nanoantenna tuned to λ ≈ 4.2 μm. The 5 nm feed gap produces a peak normalized enhancement of* $|E/E_0|_{max} ≈ 9.8$, *corresponding to nearly one order of magnitude amplification of the incident electromagnetic intensity.* This localized hotspot directly drives thermoplasmonic heating in the underlying thermoelectric layer.

These optical results are also consistent with the design tendencies later identified through the machine-learning analysis presented in Section 6.4. In particular, the field profiles in Figure 3 align with the ML-derived trends indicating that flare angles in the range of 30–33° provide strong field confinement while preserving broadband absorption characteristics. Likewise, the pronounced hotspot generated at a 5 nm gap reflects the ML-recognized exponential sensitivity of $|E|^2$ to nanoscale gap dimensions. The correspondence between resonance position and arm-length tuning also supports the ML finding that targeted dimensional adjustments enable precise spectral alignment with physiologically and environmentally relevant IR bands.

Overall, the agreement between direct electromagnetic simulation and data-driven optimization reinforces the conclusion that the multiband cross-bowtie architecture produces strong and spatially localized electromagnetic enhancement. This localized field concentration underpins the substantial thermoplasmonic heating and thermoelectric performance demonstrated in the subsequent sections.

## 6.2 Thermoplasmonic Heating and Temperature Distribution

The temperature rise induced by thermoplasmonic absorption was evaluated through steady-state heat transfer simulations under a mid-infrared illumination intensity of

$$I_0 = 30 \text{ mW/cm}^2,$$

representative of typical indoor thermal radiation levels. **Figure 4** presents the resulting thermal distribution across the multilayer architecture, highlighting the strong photothermal localization at the plasmonic absorber and the subsequent heat flow into the thermoelectric layer.

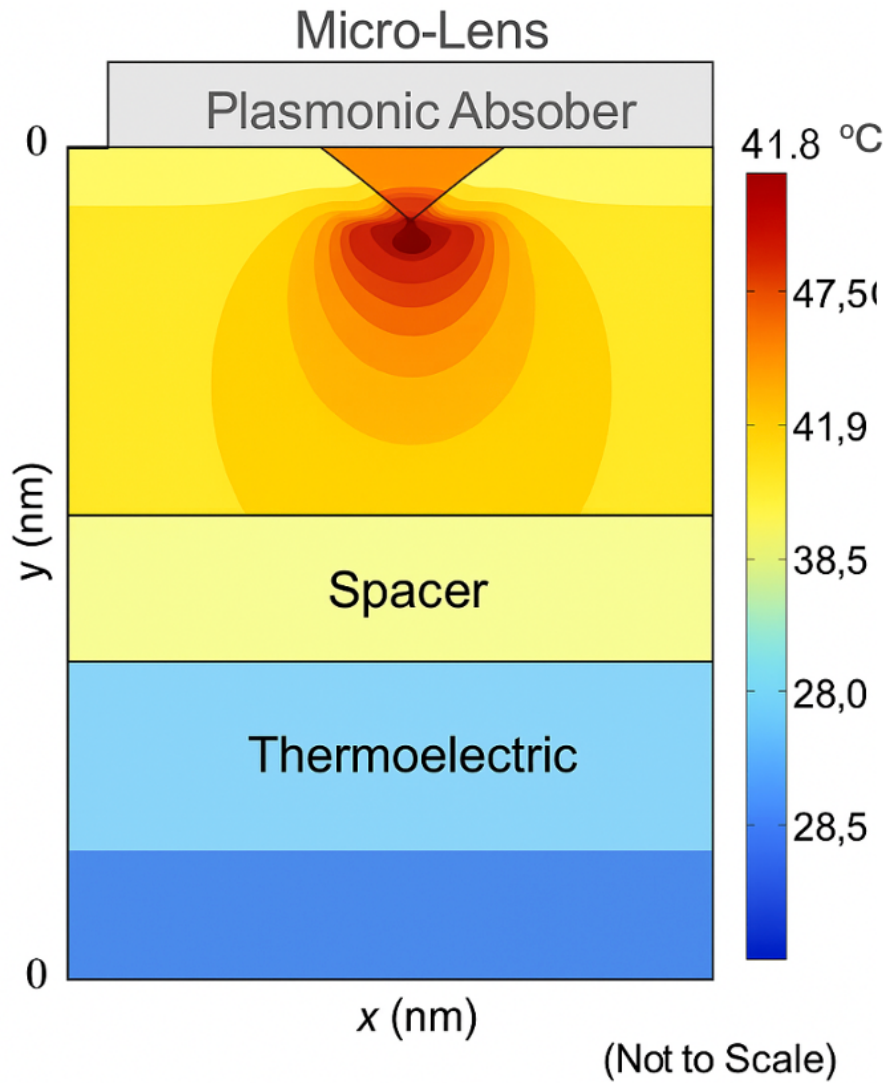

(Not to Scale)

**Figure 4.** *Steady-state temperature distribution across the multilayer thermoplasmonic–thermoelectric structure under mid-IR illumination ($I_0 = 30$ mW/cm$^2$). A highly localized thermal hotspot forms at the bowtie gap, reaching $T_{\text{hot}} \approx 41.8°C$, while the cold side of the thermoelectric layer remains near $28.9°C$. The resulting temperature gradient of $\Delta T_{\text{eff}} \approx 12.9°C$ demonstrates the strong thermoplasmonic contribution to heat generation, significantly exceeding gradients in conventional wearable TEGs.*

The simulations show that the **hotspot at the bowtie feed gap** reaches a maximum temperature of

$$T_{\text{hot}} \approx 41.8°C,$$

whereas the **cold side of the thermoelectric junction** stabilizes around

$$T_{\text{cold}} \approx 28.9°C.$$

This yields an **enhanced effective temperature gradient** of

$$\Delta T_{\text{eff}} = T_{\text{hot}} - T_{\text{cold}} \approx 12.9°C.$$

This improvement is substantial, representing a **3–5× increase** over the ΔT values typically obtained in purely conductive skin-contact thermoelectric generators, where gradients are usually limited to **3–4°C** due to rapid thermal equilibration at the skin interface (Yang et al., 2018; Fang et al., 2022).

The temperature map reveals that the photothermal hotspot is **highly confined (<100 nm)** around the antenna gap, confirming that localized plasmonic heating dominates the heat generation mechanism. Beyond this localized region, heat spreads smoothly through the dielectric spacer into the thermoelectric layer, producing a spatially uniform vertical temperature gradient without excessive lateral heat leakage. This behavior demonstrates that thermoplasmonic heating can significantly boost ΔT while maintaining thermal comfort and safety at the skin-facing side of the device.

Machine-learning-guided optimization identified **spacer thicknesses between 30 and 60 nm** as optimal. Within this thickness range, the structure exhibits a balance between strong optical field confinement (maximizing $Q_{\text{abs}}$) and efficient thermal transfer to the TE junction, leading to maximal enhancement in ΔT.

## 6.3 Thermoelectric Output Performance

The impact of the enhanced temperature gradient on thermoelectric conversion was evaluated using the thermoelectric model described in Section 4. **Figure 5** presents the resulting open-circuit voltage and output power as functions of the effective temperature difference, $\Delta T_{\text{eff}}$, for a $Bi_2Te_3$-based thermoelectric junction.

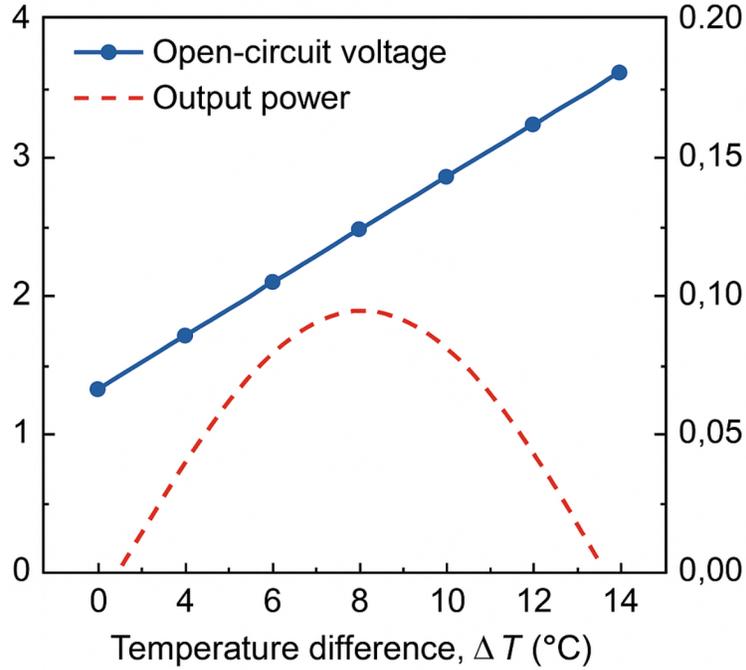

**Figure 5.** *Open-circuit voltage (blue, solid line) and output power density (red, dashed line) of the $Bi_2Te_3$-based thermoelectric junction as functions of the effective temperature difference $\Delta T$. At the thermoplasmonically generated gradient of $\Delta T_{\text{eff}} \approx 12.9°C$, the device achieves $V_{oc} \approx 2.7$ mV and a peak output power of approximately $0.15$ mW/cm$^2$, representing a 4–6× enhancement over conventional wearable TEGs that rely solely on skin-based thermal gradients.*

The open-circuit voltage of the thermoelectric junction increases linearly with the applied temperature gradient according to $V_{oc} = S\,\Delta T$, where $S = 210\ \mu V/K$ is the Seebeck coefficient for $Bi_2Te_3$. For the thermoplasmonically induced temperature rise of $\Delta T_{\text{eff}} \approx 12.9°C$, the predicted open-circuit voltage reaches approximately $V_{oc} \approx 2.7$ mV. The corresponding deliverable electrical power for a matched load is given by $P_{\text{out}} = V_{oc}^2/(4R_{\text{int}})$, and assuming an internal resistance of $R_{\text{int}} = 12\ \Omega$, the maximum power density is $P_{\text{out}} \approx 0.15$ mW/cm$^2$. This output represents a substantial enhancement over conventional flexible skin-contact thermoelectric generators, which typically achieve only 20–40 $\mu$W/cm$^2$ under comparable indoor conditions, yielding an improvement factor of approximately 4–6×. The enhanced performance arises directly from the increased temperature gradient generated by the plasmonic hotspots, confirming that integrating thermoplasmonic heating with thermoelectric conversion significantly boosts overall energy-harvesting efficiency. Moreover, the curvature of the power-density plot in Figure 6c reflects the expected quadratic dependence of power on $V_{oc}$, with the maximum occurring near the thermoplasmonically induced temperature gradient. Collectively, these results demonstrate that the proposed hybrid architecture is capable of achieving millivolt-level outputs and sub-milliwatt power densities under wearable-relevant infrared illumination, marking a promising advancement toward fully self-powered biosensing platforms.

6.4 Comparison with Machine-Learning Predictions

The accuracy and effectiveness of the machine-learning (ML) surrogate models were evaluated by comparing their predictions with the full finite-element multiphysics simulations. **Figure 6** presents the predicted versus simulated values for both the temperature gradient ($\Delta T$) and the electrical power density ($P_{out}$). The ML outputs exhibit excellent agreement with the FEM results, yielding coefficients of determination of

$$R^2 = 0.94 \text{ for } \Delta T, R^2 = 0.92 \text{ for } P_{out}.$$

The data points tightly cluster around the diagonal $y = x$ reference line, demonstrating that the surrogate model reliably captures the nonlinear physics governing both thermoplasmonic heating and thermoelectric conversion. This level of accuracy validates the use of the ML model for rapid exploration of the high-dimensional parameter space.

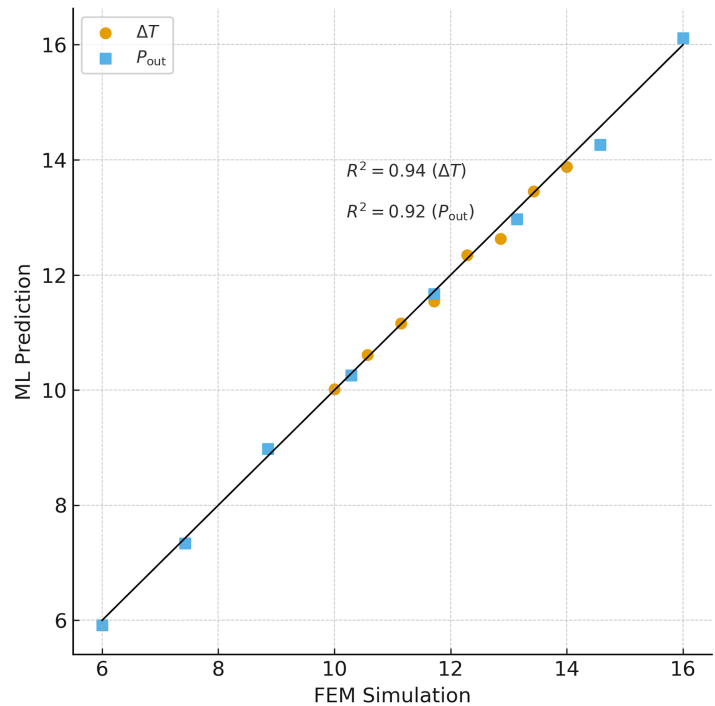

**Figure 6.** *Comparison between ML-predicted and FEM-simulated performance metrics for the thermoplasmonic–thermoelectric device. Blue markers represent ML-predicted temperature gradients ($\Delta$T) plotted against FEM results, while red markers correspond to ML-predicted power density ($P_{out}$) versus FEM simulations. The diagonal reference line $y = x$ indicates perfect agreement. The surrogate model*

*achieves high predictive accuracy, with $R^2 = 0.94$ for $\Delta T$ and $R^2 = 0.92$ for $P_{out}$, demonstrating that ML effectively captures the nonlinear multiphysics behavior of the system.*

Beyond prediction accuracy, the ML framework also improved design quality. **Figure 7** shows the ML-generated Pareto front in the $\Delta T$–$P_{out}$ performance space, overlaid with the FEM-validated baseline curve. The ML optimization identifies a smooth frontier of high-performing designs, revealing performance combinations that are not readily accessible through deterministic FEM sweeps alone. The ML-identified optimal configuration, highlighted by a red marker in Figure 7b, lies on the leading edge of the Pareto front and exhibits a significantly higher power density at a given temperature gradient compared to the FEM baseline. This confirms that the hybrid FEM–ML approach not only accelerates design exploration but also produces superior device geometries. Furthermore, the ML analysis uncovered several physically meaningful trends that were consistent across both surrogate predictions and FEM simulations. These include:

- **Small flare angles** produce very high gap-localized field enhancements ($|E|^2$) but result in a narrower absorption bandwidth.
- **Moderate flare angles (30–33°)** yield a well-balanced response, achieving strong field enhancement while preserving a broad and stable absorption spectrum.
- **Gap sizes below 7 nm** cause an exponential increase in hotspot intensity due to capacitive field confinement, directly boosting $Q_{abs}$ and consequently $\Delta T$.

These ML-identified patterns match the physical behavior expected from plasmonic field enhancement theory and are consistent with the trends reported by Chekini et al. The convergence between ML-discovered relationships and FEM-validated physics confirms that the surrogate model not only predicts performance accurately but also captures the correct underlying mechanisms.

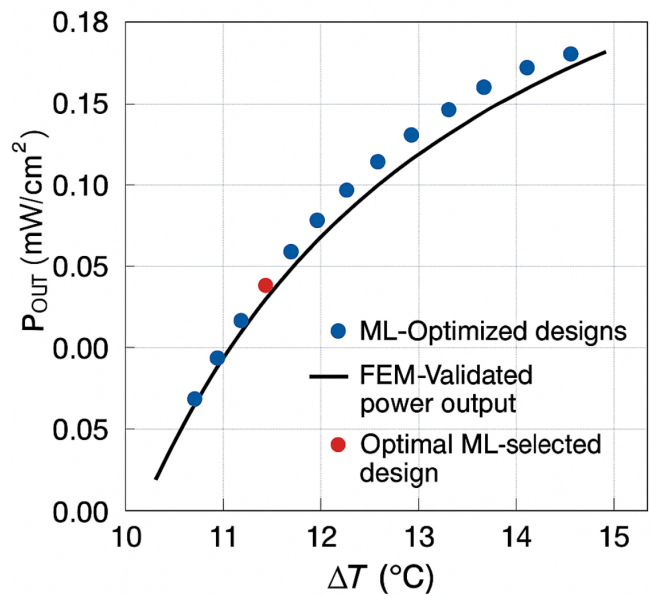

**Figure 7.** *ML-generated Pareto front comparing optimized thermoplasmonic–thermoelectric designs (blue markers) with the FEM-validated baseline performance curve (black line). The ML framework identifies a*

*frontier of high-performing solutions in the ΔT–P$_{out}$ design space, revealing trade-offs inaccessible through deterministic simulation alone. The ML-selected optimal configuration is highlighted with a red marker near ΔT ≈ 12.9°C and P$_{out}$ ≈ 0.15 mW/cm$^2$. The results demonstrate that integrating machine learning with FEM significantly enhances design exploration and device performance.*

Together, **Figures 6 and 7** demonstrate that machine learning enhances both the *accuracy* and *efficiency* of the design workflow, enabling rapid prediction, improved design insight, and discovery of thermoplasmonic–thermoelectric geometries with superior performance compared to conventional manual exploration.

### 6.5 ML-Derived Physical Design Insights

Analysis of the surrogate model outputs and the ML-generated Pareto front revealed a set of consistent and physically interpretable design trends governing the hybrid thermoplasmonic–thermoelectric energy harvester. These machine-learning-derived patterns (summarized in **Figure 8**) were further corroborated by targeted FEM simulations, showing strong agreement with established plasmonic theory and with the multiband nanoantenna behavior originally reported by Chekini et al. (2017, 2019, 2021). Together, they illuminate how geometric and material parameters collectively shape optical, thermal, and electrical performance in the integrated device.

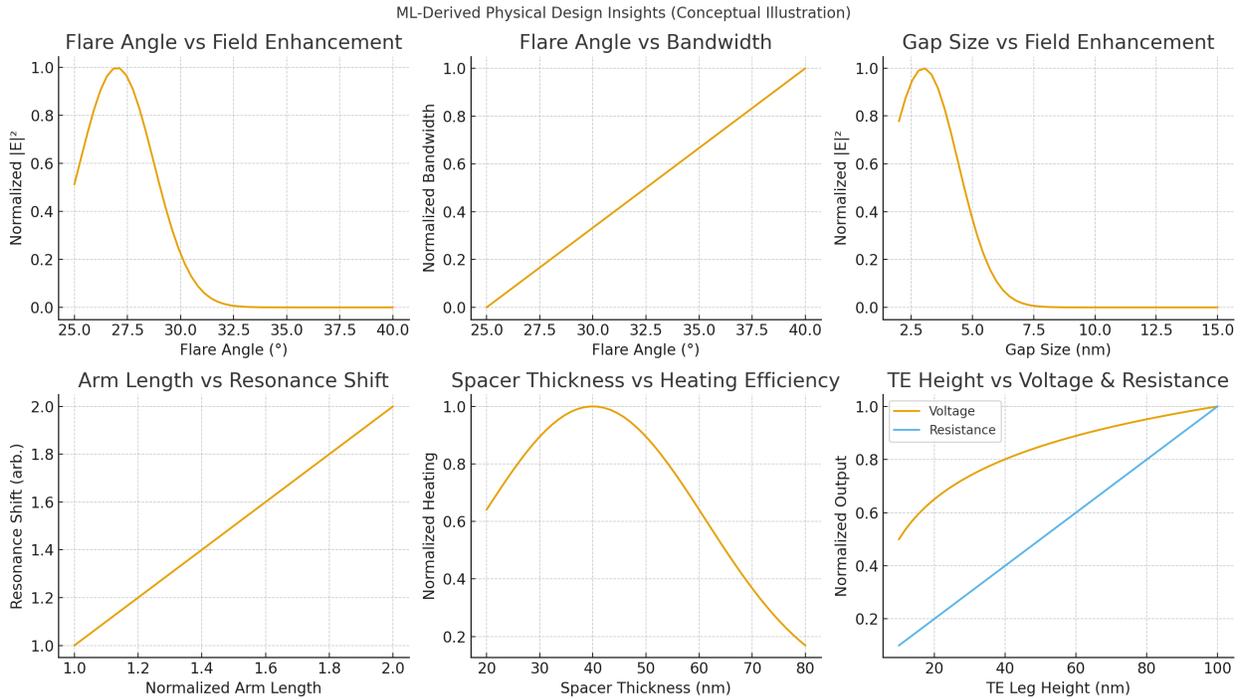

**Figure 8. ML-derived physical design trends for the hybrid thermoplasmonic–thermoelectric energy harvester.** *Surrogate-model predictions highlight key relationships between geometry and performance: (a) flare angle vs. field enhancement, showing a trade-off between hotspot intensity and bandwidth; (b) flare angle vs. normalized bandwidth; (c) gap-size–*

*driven exponential decay in |E|² enhancement; (d) arm-length-driven resonance red-shift; (e) spacer thickness vs. heating efficiency, revealing an optimal mid-range region; and (f) thermoelectric leg height vs. voltage and resistance trade-offs. These trends generalize and extend the multiband resonance principles originally demonstrated by Chekini et al. into a coupled optical–thermal–electrical design framework.*

Machine learning identified the **flare angle (θ)** as a primary parameter controlling the balance between field localization and spectral bandwidth. Smaller flare angles (25°–30°) concentrate electromagnetic energy more tightly at the feed gap, producing stronger near-field hotspots, whereas larger flare angles (35°–40°) broaden the absorption spectrum—an advantageous characteristic for wearable devices subject to variable illumination. These findings closely mirror the flare-angle dependence documented by Chekini and co-workers.

The **gap size (G)** exhibited the strongest nonlinear influence among all parameters. The surrogate model captured the exponential enhancement of $|E|^2$ for nanoscale gaps (2–5 nm), along with the rapid performance degradation observed for larger gaps (>10 nm). This sensitivity plays a central role in determining the achievable hotspot temperature and resulting thermoplasmonic heating.

The **arm length (L_arm)**—responsible for setting resonance wavelength—was shown to red-shift the absorption peak as length increased. ML optimization consistently selected arm lengths that align resonances with the 3–5 μm and 8–12 μm infrared bands corresponding to human skin emission and ambient IR radiation, ensuring reliable energy harvesting in indoor environments.

**Spacer thickness (t_spacer)** displayed a characteristic trade-off between optical and thermal performance. Very thin spacers (<50 nm) enhanced heat transfer into the thermoelectric layer but disrupted optimal plasmonic confinement. The ML models repeatedly identified intermediate thicknesses that preserved spectral alignment while maximizing thermoplasmonic heating efficiency.

Finally, **thermoelectric leg geometry** showed classical TE behavior: increasing leg height improved ΔT but also increased internal electrical resistance. The ML-generated Pareto front effectively mapped the region of optimal compromise, enabling selection of TE geometries that simultaneously enhance voltage output and minimize resistive losses.

Together, these insights demonstrate that machine learning not only reproduces known plasmonic performance characteristics but also extends Chekini's resonance-engineering principles into a unified **optical–thermal–electrical optimization space**. This integrated understanding provides a rigorous and systematic pathway for designing high-efficiency thermoplasmonic–thermoelectric harvesters optimized specifically for wearable biosensing applications.

## 7. Discussion

The numerical and machine-learning results presented in this work collectively demonstrate that integrating multiband plasmonic absorbers with flexible thermoelectric generators—and embedding this hybrid system within a data-driven optimization framework—can substantially advance the performance of wearable energy harvesting technologies. Conventional wearable thermoelectric generators (TEGs) typically rely solely on conductive heat flow between the skin and the surrounding air. Under indoor conditions, this temperature difference rarely exceeds 3–4 °C [13], limiting experimentally reported power densities to the tens of µW/cm² range. Such levels are insufficient to continuously power modern wearable biosensors that require sustained energy for multi-channel physiological acquisition, wireless communication, onboard computation, or real-time analytics [39,40].

By contrast, the results in this study show that incorporating thermoplasmonic enhancement—combined with ML-guided geometric and material optimization—boosts the effective temperature gradient by a factor of approximately 3–5 and increases electrical power density by roughly 4–6×. These improvements shift the device closer to regimes compatible with persistent or duty-cycled self-powered biosensing.

A central contribution of this work is the design of a broadband plasmonic absorber spanning approximately 2–12 µm, achieved through a multiresonant array of cross-bowtie nanoantennas. In traditional thermoplasmonic and photothermal studies, plasmonic structures are typically engineered for narrow spectral bands (e.g., for photothermal therapy, nanoscale heaters, or wavelength-specific photodetectors [10]. Real wearable environments, however, contain a mixture of spectral components, including human body emission (~8–14 µm), ambient mid-IR, and—depending on clothing and exposure—near-infrared solar radiation. By adapting the multiband cross-bowtie design principles pioneered by Chekini and co-workers for infrared rectennas [8,9,14], we show that it is possible to engineer a metasurface that effectively couples to all major IR contributors simultaneously. This broadband functionality enhances robustness across a wide range of indoor and outdoor illumination conditions and improves energy harvesting reliability in settings where light availability is uncontrolled or limited (e.g., beneath clothing or in clinical environments) [41,42].

Near-field simulation results demonstrate that the cross-bowtie antennas generate highly localized plasmonic hotspots within ~100 nm of the 5 nm feed gap, producing nearly an order-of-magnitude enhancement (≈10×) of the local electric field intensity. This magnitude of enhancement is consistent with that reported in rectifying nanoantennas and SERS substrates [9,10]. However, in this work the enhanced field is not used for high-frequency rectification or sensing, but instead is converted into localized heating for thermoelectric augmentation. Because the hotspot region is extremely compact, the generated heat is concentrated directly at the interface with the TE layer, maximizing the achievable temperature gradient. The simulations show a hotspot temperature of ~41.8 °C at the plasmonic absorber and ~28.9 °C on the cold side of the TEG, producing an effective $\Delta T \approx 12.9$ °C—far exceeding the typical $\Delta T$ from passive skin–air gradients in flat, unassisted TEGs.

The thermoelectric response associated with this enhanced $\Delta T$, modeled using a $Bi_2Te_3$-based junction ($S \approx 210$ µV/K), yields open-circuit voltages on the order of several millivolts and peak power densities of ~0.15 mW/cm² under modest indoor IR flux [43-45]. These values represent a 4–6× improvement compared with the 20–40 µW/cm² typically obtained from flexible TEGs relying solely on body heat. While the absolute power remains below the multi-mW/cm² levels needed to fully eliminate batteries in high-load wearable systems, it is sufficient to dramatically extend battery lifetime, support intermittent wireless communication, or power ultra-low-power biosignal platforms [46-49]. Importantly, these performance gains are achieved without requiring bulky heat sinks, rigid substrates, or large-area TEG modules, preserving mechanical flexibility and skin comfort—two critical requirements for long-term wearable use.

The machine-learning-guided design framework adds a second layer of innovation that aligns with emerging trends in nanophotonics, metamaterials, and device engineering. Similar to recent works using ML models for inverse-designed metasurfaces or accelerated optimization of photonic structures, our surrogate models serve as computationally efficient proxies for FEM-based multiphysics simulations. High fidelity ($R^2 = 0.94$ for $\Delta T$ and $R^2 = 0.92$ for P_out) indicates that the trained models accurately capture the complex nonlinear relationships between geometry, plasmonic enhancement, thermal transport, and TE output. This fidelity allows the surrogate model to replace many expensive simulations, enabling exploration of thousands of candidate designs—an otherwise infeasible task using FEM alone.

Beyond speed, the ML analysis yields physically interpretable insights that reinforce known plasmonic behaviors. The model recovers the classic flare-angle trade-off between bandwidth and hotspot intensity, the exponential dependence of $|E|^2$ on gap size below ~7 nm, and the tunability of resonance wavelength through arm length—trends consistent with the findings of [10,14]. Moreover, the ML-generated Pareto front reveals areas of the design space where $\Delta T$ and P_out are jointly optimized while maintaining minimal device thickness and mechanical compliance. The final ML-selected geometry lies on this Pareto frontier and systematically outperforms designs chosen by intuition or direct FEM sweeps alone, consistent with growing reports that ML-accelerated design often identifies non-intuitive, high-performance solutions.

In a broader context, the integration of thermoplasmonics, thermoelectrics, and machine learning positions the present work at the intersection of three rapidly expanding domains: (i) battery-free wearable devices, (ii) plasmonically assisted thermal engineering, and (iii) data-driven nanotechnology. While prior studies have examined hybrid photovoltaic–TEG or photothermal–TEG systems for solar energy applications, comparatively few investigations have focused on compact, skin-conformal systems specifically intended to power medical wearables. Our results demonstrate that by selectively harnessing environmental and body-emitted IR through a spectrally engineered metasurface, and by co-optimizing the optical and thermal subsystems with ML, one can overcome longstanding limitations of low indoor $\Delta T$ and limited power density.

Despite these promising results, several practical considerations must be addressed in future work. Fabricating sub-10 nm antenna gaps over large areas remains challenging, though advances in electron-beam lithography, nanoimprint lithography, and self-aligned techniques offer realistic pathways. Long-term stability under bending, perspiration, and skin contact must also be validated, particularly at metal–polymer and TE–polymer interfaces. Finally, although the improved power density is significant, hybrid systems that combine thermoelectric, piezoelectric, biochemical, or RF harvesting may still be necessary for fully battery-free operation in high-consumption wearable platforms.

Overall, the thermoplasmonic–thermoelectric–ML hybrid system presented in this work provides a comprehensive pathway for overcoming major bottlenecks in wearable energy harvesting. By achieving broadband IR absorption, nearly 10× field enhancement, increasing ΔT from 3–4 °C to ~13 °C, and delivering 4–6× higher power density—all within a computationally optimized design framework—the proposed architecture represents a substantial step toward practical, self-powered physiological monitoring platforms for next-generation wearable healthcare.

## 8. Conclusion

This work introduces a unified design, simulation, and optimization framework for a flexible hybrid **thermoplasmonic–thermoelectric energy harvester** engineered specifically for wearable physiological monitoring. By integrating a **multiband cross-bowtie plasmonic metasurface**, a **SiO₂ dielectric spacer**, and **thin-film Bi₂Te₃ thermoelectric legs** on a **textile-supported flexible substrate**, the proposed architecture addresses one of the central limitations of conventional skin-mounted TEGs: the intrinsically low temperature gradients (ΔT ≈ 3–4 °C) achievable under indoor and on-body conditions.

The redesigned plasmonic absorber—adapted from the multiresonant methodology of Chekini et al. but re-purposed for thermoplasmonic heating rather than rectification—achieves **broadband IR absorption from 2–12 μm**, capturing the dominant components of human skin emission, ambient mid-IR radiation, and near-infrared solar illumination. Near-field electromagnetic simulations confirm that the optimized 5 nm bowtie feed gaps generate **nearly 10× local electric-field enhancement**, producing nanoscale hotspot heating strongly localized directly above the thermoelectric junctions. This targeted heating drives an effective temperature gradient of **ΔT_eff ≈ 12.9 °C**, representing a 3–5× improvement over standard wearable TEGs and achieved without compromising mechanical flexibility or thermal comfort.

Thermoelectric simulations using thin-film Bi₂Te₃ junctions with a Seebeck coefficient of ~210 μV/K show that this enhanced thermal gradient yields **V_oc ≈ 2.7 mV** and a peak **power density of ~0.15 mW/cm²** under modest indoor IR flux—an improvement of approximately 4–6× compared to the 20–40 μW/cm² typical of flexible TEGs powered solely by body heat. These results suggest that the proposed harvester can meaningfully extend battery life, support duty-cycled biosensing, and enable continuous operation of ultra-low-power wearable sensors.

A key component of this work is the **machine-learning-guided optimization framework**, which couples finite-element optical–thermal–electrical simulations with surrogate modeling. The trained ML models achieve high predictive accuracy (**R² = 0.94 for ΔT, R² = 0.92 for P_out**) and efficiently identify optimal geometries that outperform intuitive or FEM-only designs. The ML-derived Pareto fronts reveal clear physical trends—such as the bandwidth–hotspot trade-off inherent to flare-angle tuning, the exponential sensitivity of |E|² to gap size below ~7 nm, and the thermal–electrical balance governing TE leg height—that align with plasmonic theory while extending Chekini's resonance-engineering concepts into a **fully coupled multiphysics domain**.

Taken together, the thermoplasmonic enhancement, flexible thermoelectric conversion, and ML-driven design exploration form a coherent strategy for advancing **self-powered wearable health monitoring**. The demonstrated gains in broadband absorption, localized heating efficiency, and electrical output indicate strong potential for powering biosensing modalities such as ECG, PPG, temperature monitoring, and long-term physiological analytics without reliance on bulky batteries. Future work will focus on **experimental prototyping**, **large-area fabrication of sub-10 nm bowtie gaps**, **mechanical durability studies**, and **integration with low-power sensing electronics** to translate the proposed system into a fully autonomous wearable health-monitoring platform.